\documentclass[conference]{IEEEtran}
\usepackage{xcolor}
\usepackage[pdftex]{graphicx}
\usepackage{mathtools}
\begin{document}
%
\title{Ontology-based Adaptive e-Textbook Platform for Student and Machine Co-Learning}



%
\author{\IEEEauthorblockN{Noel Nuo Wi Tay\IEEEauthorrefmark{1},
Sheng-Chi Yang\IEEEauthorrefmark{1},
Chang-Shing Lee\IEEEauthorrefmark{1}\IEEEauthorrefmark{2} and
Naoyuki Kubota\IEEEauthorrefmark{3}}
\IEEEauthorblockA{\IEEEauthorrefmark{1}Center for Research of Knowledge Application and Web Service, National University of Tainan, Taiwan. \\
Email: taynoel@ieee.org, sc.yang.mark@gmail.com}
\IEEEauthorblockA{\IEEEauthorrefmark{2}Dept. of Computer Science and Information Engineering, National University of Tainan, Taiwan\\
Email: leecs@mail.nutn.edu.tw}
\IEEEauthorblockA{\IEEEauthorrefmark{3}Graduate School of System Design, Tokyo Metropolitan University, Tokyo, Japan\\
Email: kubota@tmu.ac.jp}}


\maketitle

\begin{abstract}
The use of electronic textbooks (e-book) has been heavily studied over the years due to their flexibility, accessibility, interactivity and extensibility. Yet current shortcomings of e-book, which is often just a digitized version of the original book, does not encourage  adoption. Consequently, this leads to a rethinking of e-book that should incorporate current technologies to augment its capabilities, where inclusion of information search and organization tools have shown to be favorable. This paper is on a preliminary work to add intelligence into such tools in terms of information retrieval. Construction of knowledge graph for e-book material with little overhead is first introduced. Information retrieval through typed similarity query is then performed via random walk. Case study demonstrate the applicability of the e-book platform, with promising application and advancement in the area of electronic textbooks.
\end{abstract}

\IEEEpeerreviewmaketitle

\section{Introduction}


The use of e-books has been heavily studied over the years \cite{liang2015effects,chan2015exploration}. Their use in education has been anticipated due to their flexibility, accessibility, interactivity and extensibility \cite{daniel2013textbooks,murray2011textbooks}. From teacher's perspective, e-books also prove useful in monitoring\cite{embong2012teachers} and for student progress analysis\cite{JUNCO201554}. Through this platform, teachers can create and present their teaching materials for the students to access, whereas for the students, they can use the platform to learn, communicate, take notes, do pre-study and so on. E-books have also been extended to become a platform for e-learning and e-publishing \cite{gu2015design}, as well as learning material recommendation \cite{lee2015fml}.    

As of now, a widely recognized approach to develop e-books is to digitalize the existing printed books as replacement textbooks \cite{gu2015design}. Yet, such a system will be useless if all it does is to mimic the physical book digitally. Currently, majority of e-book publishers are still producing such simple digital versions of the original printed books, yet at the same time, there is no sense of familiarity with it \cite{gu2015design}, which leads to failure for endeavours that capitalizes on it \cite{kim2010development}. Besides, studies still show that printed books are still preferred\cite{woody2010books}, and that e-books are often being treated as reference books \cite{abdullah2008students}. Consequently, this leads to a rethinking of what e-book is, and how technologies can be incorporated with it \cite{kim2010xml}. This is because people treat e-books differently, such as preferring short texts \cite{brunet2011incoming}, and also a preference for hyperlinks instead of page flipping\cite{chong2009design}. Functionalities unique to the e-book such as search and navigation functionalities have been receiving favorable responses\cite{brunet2011incoming,mar2009textbook}, as well as annotation and sharing capabilities that support learning\cite{lim2014students}. Recent studies have shown that text, highlighting, bookmarking, multimedia, translation, dictionary and encyclopedia tools are popular components that need to be taken into account in the development of e-book and/or its supporting platform\cite{sheen2015relationship,Sheen2017}. 

As most tools are geared towards search and information organization, and intelligent system that supports the management of information according to how the user interacts with the e-book (such as annotation, bookmarking and highlighting) not only simplifies such interactions, but also augments the ability of the human reader in accessing and managing ever more information during the course of learning. But, intelligent platforms come at a price where the teacher needs to painstakingly create their teaching material, which can be a huge overhead. Authoring tools are required to be easy to use for wide adoption of the technology \cite{gu2015design}.

This work strives to handle the task of developing an intelligent platform that not only reduces the overhead for teaching material creation in the e-book, but that it can be a personalized information management system for both the teacher and student. This can also induce a more collaborative learning, where studies had found to be favorable\cite{mcfall2006experiences}. Teaching materials like books, slides and notes are normally presented in a way that assist human learners to garner knowledge that enables continual learning, where earlier knowledge support the assimilation of more complicated concepts further up the course. Therefore, these learning materials normally have internal structures that associates the represented knowledge. For example, descriptions to explain a particular topic on computer number systems will have an order, such as the explanation description on binary-decimal conversion will be preceded by an explanation description on what binary number system is. Another example is the topic classes that categorizes the descriptions. This shows that teaching materials can be represented as directed graphs that connects various nodes, where these nodes themselves may contain values (such as a chunk of text). Such graphs for learning purposes have been studied, such as in language study \cite{lee2016fml}. This paper will denote this as knowledge graph. There are a lot of works that utilizes such knowledge graphs for information retrieval purposes  such as query expansion\cite{collins2005query}, in natural language processing \cite{toutanova2004learning}, gene identification \cite{arnold2009information,Cohen2006}, social networks \cite{liben2007link}, probabilistic programming \cite{yang2017differentiable} and large scale knowledge-base inference \cite{lao2011random}. It has been shown that information retrieval via graphs outperforms that of texture similarities \cite{minkov2006contextual, Cohen2006}. 

The aim of this paper is to present a preliminary work on the platform's development, where emphasis is given on how teaching materials are to be created and how queries pertaining to learning information management can be achieved. Such queries can be associated with the tools in the e-book such as bookmarking and highlighting. With the use of OWL DL, knowledge graph can be easily constructed by the teacher. The generated knowledge graph can also be used as a Markov Chain for querying purposes. For clarity, domain experts are people who will construct the book ontology. In this paper, ``teacher" will be used interchangeably with ``domain experts". Case study is on the query between questions and descriptions of the teaching material. This paper does not assume the use of any prior knowledge base.

This paper is organized as follows: Section \ref{sec:platformArchitecture} describes the overall architecture of how the e-book platform is set up and deployed. Then, Section \ref{sec:teachingMaterialGraph} explains how the knowledge base can be set up, as well as the generation of additional facts. Given the generated knowledge graph, Section \ref{sec:graphWalk} explains how it can be used for information retrieval via graph walk. The platform is then evaluated in Section \ref{sec:experiment}. Final, conclusion is given in Section \ref{sec:conclusion}.

\section{Platform Architecture}
\label{sec:platformArchitecture}
\begin{figure}
	\begin{center}
		\includegraphics[angle=0, width=1\columnwidth]{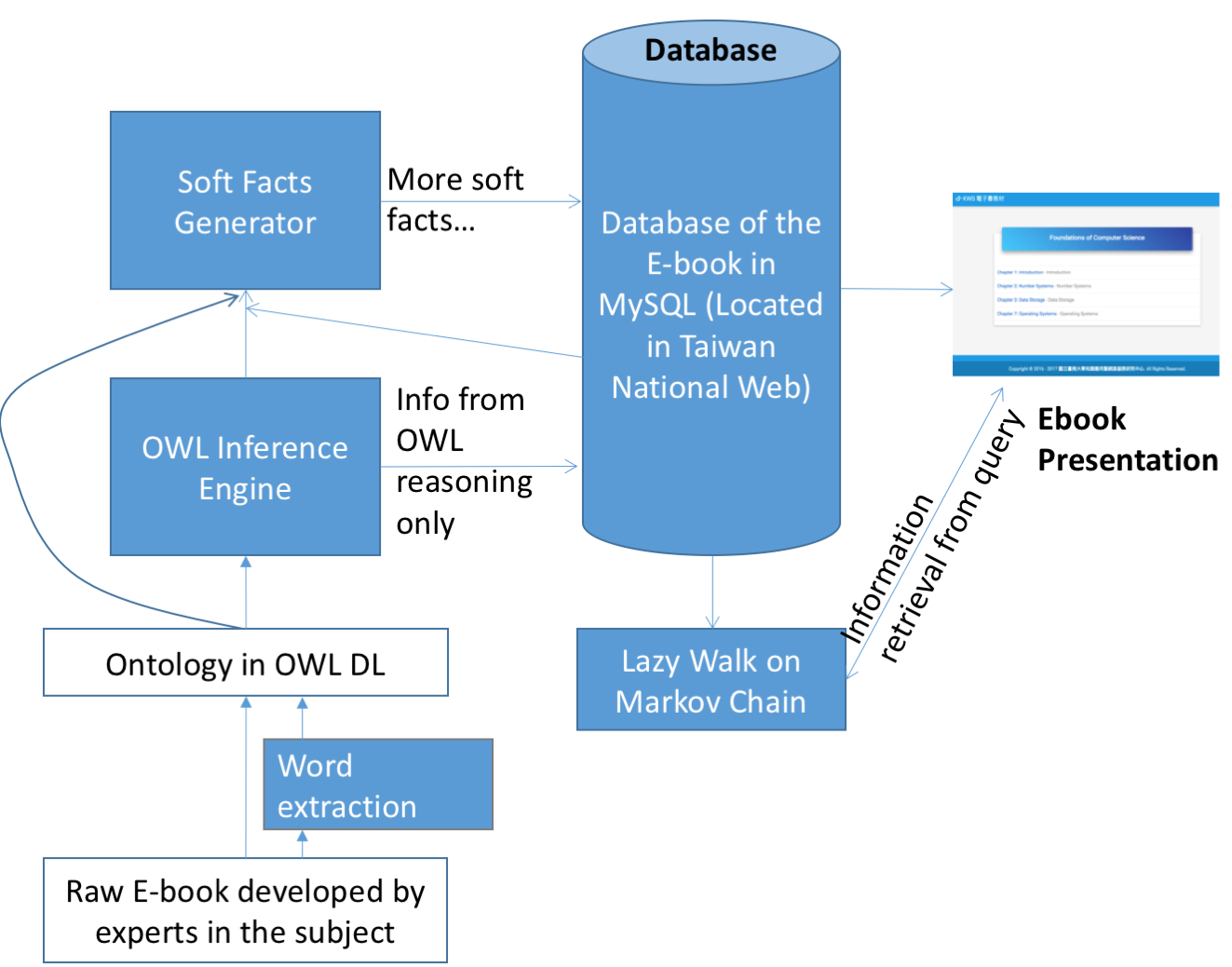}
	\end{center}
	\caption{Architecture of the Electronic Textbook Platform}
	\label{fig:architecture} 
\end{figure}

Figure \ref{fig:architecture} shows the architecture that implements the electronic book platform. Explanation of the architecture will be given according to process flow. It is assumed that the teacher knows everything about the flow of the teaching material and the topic hierarchy. This architecture will leverage this knowledge to build additional knowledge such that more intelligent tasks can be achieved.

\begin{figure}
	\begin{center}
		\includegraphics[angle=0, width=1\columnwidth]{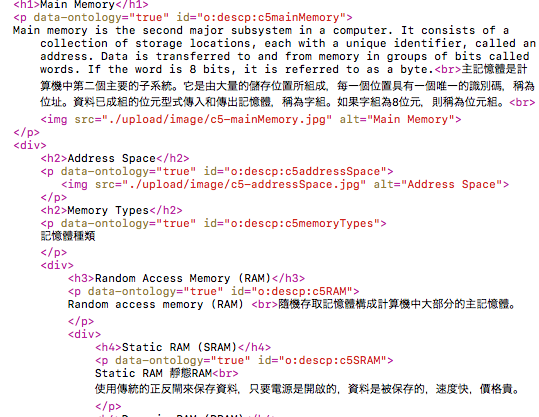}
	\end{center}
	\caption{Example of a part of Teaching Material presented in HTML}
	\label{fig:htmlExample} 
\end{figure}

\begin{figure}
	\begin{center}
		\includegraphics[angle=0, width=1\columnwidth]{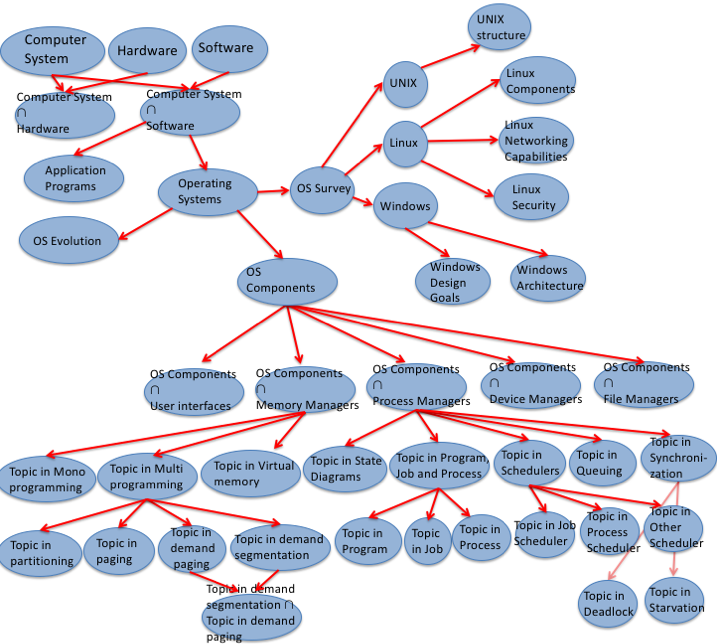}
	\end{center}
	\caption{Example Topic Hierarchy for the Chapter on Operating System}
	\label{fig:topicHier} 
\end{figure}

Given a course a student need to take, initially, teachers who are expert in the field need to create the teaching materials via some standards specified in Section \ref{sec:teachingMaterialGraph}. In this work, the  teaching materials are presented in HTML format, where there are annotations for the topic and description ID. Description is a small part of text (and images) in the book, which may be a whole page, a paragraph or a sentence (where the teacher has the freedom to decide).  Topic, which is often arranged in a hierarchical manner (where an example is shown in Figure \ref{fig:topicHier}), is a class that contains the description. In terms of OWL DL specification as described in Section \ref{sec:nodeDefinition}, topic will be a class object that contains the instance of description. An example teaching material presentation in HTML is shown in Figure \ref{fig:htmlExample}. The way the teaching material is written in the HTML is such that description and their respective topic can be extracted. Besides that, the topic hierarchy can be obtained as well via the headers. From Figure \ref{fig:htmlExample}, it can be observed that there is a description with id ``o:descp:c5mainMemory" that is under the topic ``main memory". Apart from that, it can also be observed that ``address space" topic is a sub-topic of ``main memory". Although not shown here, all topics and descriptions will be given unique IDs.

The extracted description instance and topic hierarchy and their association will be described via OWL DL. At this stage, only the ID is used. Every words and their association to the descriptions will also be extracted and added into the ontology. There are also additional axioms that are added into the ontology to ensure the subsequent reasoner to function properly as explained in Section \ref{sec:additionalFacts}.

The facts and ontologies will then be loaded into the OWL inference engine to generate all the facts that can be deduced, before loading all of them to the MySQL database. Although not implemented yet, there is also an additional soft facts generator that generates facts with a certain magnitude of certainty. This soft facts generator is not used in this paper as the main purpose is to study the information retrieval given the ontologies. 

Users will then access the website that shows the teaching material exactly as how the HTML is designed, and also with semantic links stored in the database. Just like web services and semantic web, a lot of features and services can be realized with such setup. 

The knowledge graph stored in the database will also be converted to a Markov Chain that will be used for information retrieval purposes. Users can access this feature via the website, which will be explained in  Section \ref{sec:graphWalk}. This information retrieval mechanism using the graph represented by OWL DL is the main emphasis of this paper.

\section{Teaching Material as a Graph}
\label{sec:teachingMaterialGraph}
\begin{figure}
	\begin{center}
		\includegraphics[angle=0, width=1\columnwidth]{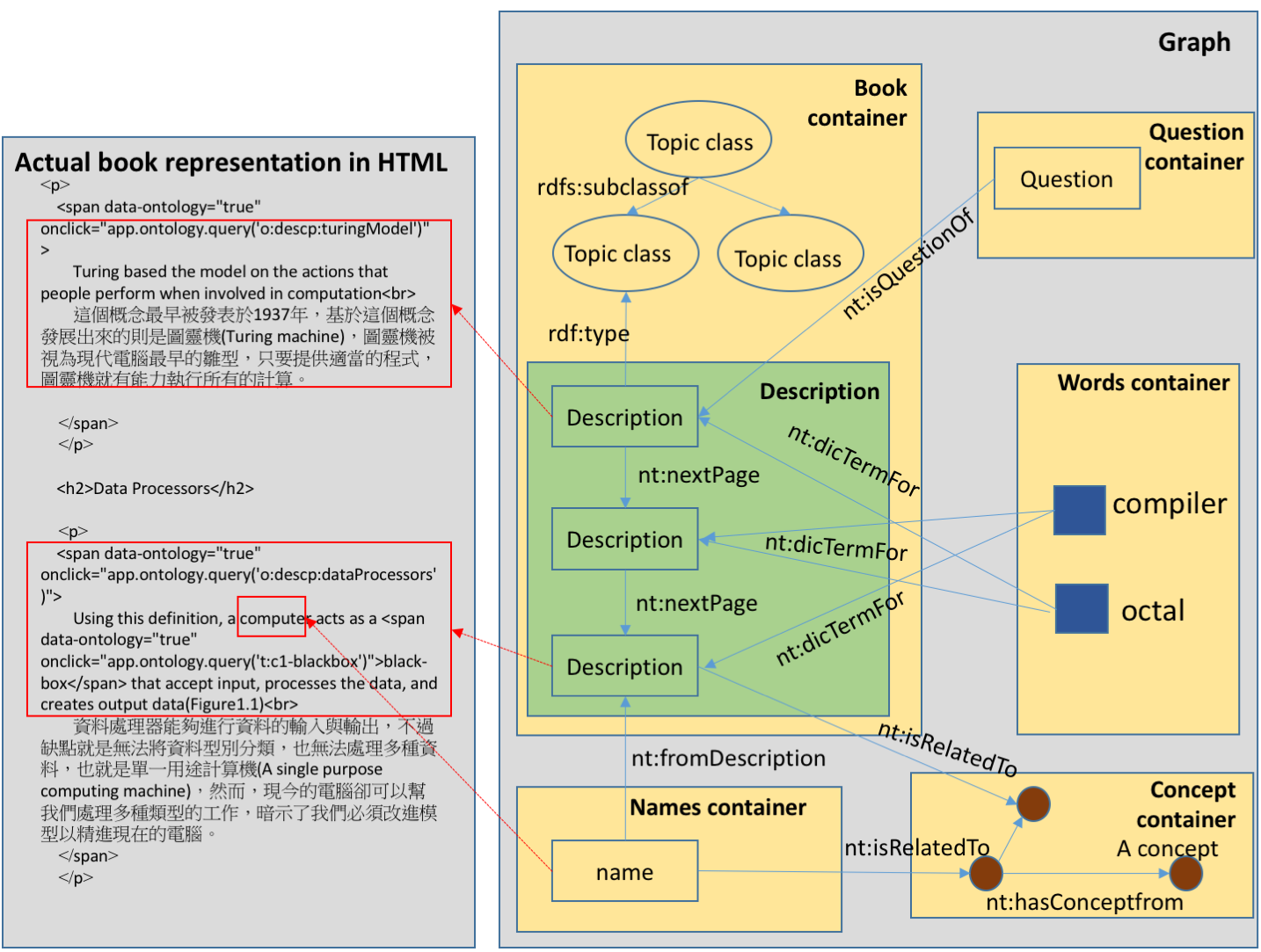}
	\end{center}
	\caption{A Simple Knowledge Graph Example for Teaching material}
	\label{fig:knowledgeGraph} 
\end{figure}

A graph $G$ consists of a set of nodes and a set of labeled directed edges between the nodes. This section will describe how the teaching material graph can be constructed. A simple teaching material knowledge graph is shown in Figure \ref{fig:knowledgeGraph}. Note that from the figure, the knowledge graph only handles the relationship of the nodes (right side of the figure). The data associated with the nodes will be accessed through the HTML representation (left side of the figure), which will be denoted as the value of the nodes. Contrary to the node itself, the value of the node will not be used for reasoning.

\subsection{Node Definition}
\label{sec:nodeDefinition} 
Teaching material knowledge graph utilizes some objects, namely, description, topic, question, name, concept and term. All except term object needs to be manually defined by the teacher. These objects are explained as follows:\\
\\
Description: As shown in Figure \ref{fig:knowledgeGraph}, description node is associated with a chunk of text/images from the HTML representation of the teaching material. Every description has a unique ID, and no nested description is allowed. Description is an instance object in the OWL DL representation under the type ``book container".\\
\\
Topic: Description is the instance, whereas topic is the class that categorizes the description based on the topic hierarchy of the material. One can think of it as the hierarchy from the table of contents of a book. Note that teachers can also categorize a description under two different topics according to how he/she seems fit. In OWL DL representation, topic is a class object, which is a sub-class of ``book container".\\
\\
Question: Question is an instance object, which is associated with a value containing the actual question. Question is linked with description that it is relevant to. Note that a question can have multiple links to different descriptions, and vice versa. Question is under the type ``question container".\\
\\  
Name: Contrary to description, which stores a chunk of text, name is an instance object that is associated with a name within the description. Note that the value of name is not limited to one word. Name is under the type ``name container".\\
\\
Concept: Concept is an instance object, which is used to encode the abstract concept pertaining a description or name. Concept object is expected to be manually designed by domain expert. Automatic generation of concept is subject of future work. Concept is under the type ``concept container".\\
\\
Term: Name object is associated with a manually annotated word/s, whereas term is automatically extracted. Term is under the type ``term container"\\
\\
Containers: The purpose of type similarity query is to return a ranked list of objects associated with a certain type given some input nodes. The containers explained previously are used as such filter for typed output.\\ 

\subsection{Initial Ontology}
\label{sec:initialOntology} 
OWL DL provides a means to generate additional hard facts given an ontology and some initial facts. One can exploit such reasoning capability to enrich the teaching material knowledge graph, without requiring teachers to specify every node links possible. The initial ontology is explained as follows:\\
\\
Class relationship between topic objects: As explained in Section \ref{sec:nodeDefinition}, topics are arranged in a hierarchical manner according to how the teaching material is structured. One can endow topics, which are of class object in OWL DL representation, with $rdfs:subClassOf$ property to generate super and sub-class relationships.\\
\\
Description relationship with topics: In OWL DL representation, description and topics are related via $rdf:type$ property, where, combining the super and sub-class relationship described previously, can generate more instance associations with topics.\\  
\\
Other properties: Subsequent descriptions are linked via the property $:nextPage$, which has an inverse $:prevPage$ via $owl:inverse$. Question is linked to description via property $:isQuestionOf$, name linked to description through property $:fromDescription$, description and name to concept through property $:isRelatedTo$, and term to description through property $:dicTermFor$, all of which have inverse properties respectively.
    
\subsection{Creating Initial Facts}
\label{sec:creatingFacts} 
The teacher is responsible to create the initial facts. To prevent the huge overhead imposed on the teacher to create teaching material, and at the same time, able to obtain ample information for reasoning, the process flow is explained in 3 procedures.

\subsubsection{Procedure 1}
Topic hierarchy should be specified, which is shown in the $Topic$ region of Figure \ref{fig:topicDescpLink}. $rdfs:subClassOf$ property is used for topic-topic links. At this stage, the teacher only considers the category and their hierarchy of the teaching material (which is just the structure in the tables of contents), and does not need to consider the actual text of the teaching material.

\subsubsection{Procedure 2}
At this stage, the teacher will separate the teaching material into chunks of description, and link them via $:nextPage$, which is shown in the $Description$ region of the Figure \ref{fig:topicDescpLink}. The teacher does not need to consider the topic category of the descriptions at this stage. At the same time, given the description, the teacher can also link question object to their relevant descriptions. During this stage, association between the text chunks in the HTML representation and the question and description object should be made as shown in Figure \ref{fig:knowledgeGraph}.

\subsubsection{Procedure 3}
At this stage, links between description and topic are made. The teacher can also link a description to multiple topics. 

\begin{figure}
	\begin{center}
		\includegraphics[angle=0, width=1\columnwidth]{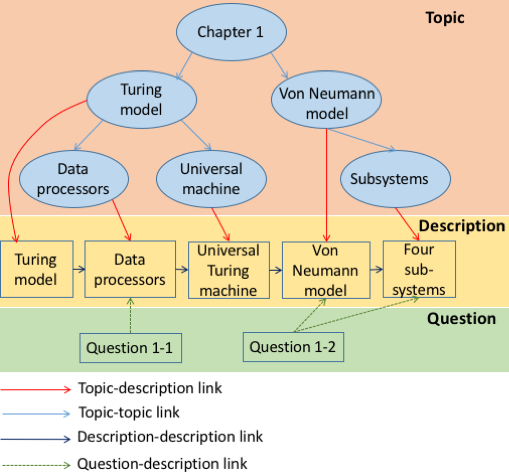}
	\end{center}
	\caption{Teaching Material Creation Example}
	\label{fig:topicDescpLink} 
\end{figure}

All three aforementioned procedures can be automatically extracted from the HTML representation of the teaching material. These procedures also pose minimal overhead as such information from teaching materials ought to be ready for any classes. Given the example shown in Figure \ref{fig:topicDescpLink}, the following shows the triples (subject, property, object) in OWL DL representation:\\
\scriptsize
$topic:Turing\_model,rdfs:subClassOf,topic:Chapter\_1$\\
$topic:Von\_Neumann\_model,rdfs:subClassOf,topic:Chapter\_1$\\
$topic:Data\_processors,rdfs:subClassOf,topic:Turing\_model$\\
$topic:Universal\_machine,rdfs:subClassOf,topic:Turing\_model$\\
$topic:Subsystems,rdfs:subClassOf,topic:Von\_Neumann\_model$\\
$dsc:Turing\_model,:nextPage,dsc:Data\_processors$\\
$dsc:Data\_processors,:nextPage,dsc:Universalturingmachine$\\
$dsc:Universalturingmachine,:nextPage,dsc:VonNeumannmodel$\\
$dsc:VonNeumannmodel,:nextPage,dsc:FourSubsystems$\\
$dsc:Turing\_model,rdf:type,topic:Turing\_model$\\
$dsc:Data\_processors,rdf:type,topic:Data\_processors$\\
$dsc:Universalturingmachine,rdf:type,topic:Universal\_machine$\\
$dsc:VonNeumannmodel,rdf:type,topic:Von\_Neumann\_model$\\
$dsc:FourSubsystems,rdf:type,topic:Subsystems$\\
\normalsize

\subsection{Additional Facts Generation}
\label{sec:additionalFacts} 
Given the ontology and facts provided in Section \ref{sec:initialOntology} and \ref{sec:creatingFacts}, additional facts can be generated. Yet, the accumulation of these facts might not be enough as they are quite sparse considering the number of nodes. One way to increase connection is to have domain expert add in concept nodes and their relationships as described in Section \ref{sec:nodeDefinition}, but such task will be extremely tedious and does not scale well. 

In this work, words are automatically extracted from the descriptions, where these words will be associated with their respective descriptions, which provides a denser relationship between the nodes.

\section{Typed Similarity via Graph Walk}
\label{sec:graphWalk}
Similarity evaluation between two nodes is performed using random walk, which is explained in this section. 

The generated knowledge graph explained in Section \ref{sec:teachingMaterialGraph} is converted to a Markov Chain, taking into account all nodes and properties, including their inverses. Note that inverses for $rdf:type$ and $rdfs:subClassOf$ are also generated. Similarity query is likened to traversing from some starting input nodes, where the nodes that most likely the random walk will end up in is the most similar item (output), given that correct node type is selected. This paper will employ the random walk employed in \cite{Cohen2006}.

Given a node, $x$, in the Markov Chain, to walk away from the node, one needs to first choose the edge (property) type $\ell$ to move out from. Then, the next node that the chosen edge type is leading to is randomly chosen. Lets denote $x\xrightarrow{\ell}y$ the directed edge $\ell$ going from node $x$ to node $y$. Given $L(x)=\{\ell:\exists y \text{ } x\xrightarrow{\ell}y\}$, then, the probability of choosing $\ell$ is uniform over all label types that extended out of $x$, which is:\\
\begin{equation}
\label{eq:chooseL}
P(\ell|x)=\frac{1}{|L(x)|}
\end{equation} 

$y$ is then uniformly chosen given the edge type $\ell$. Let's define the set of nodes that extends from $x$ given $\ell$ as $Y(x,\ell)=\{y:x\xrightarrow{\ell}y \}$. Thus, the probability of choosing node $y$ given node $x$ and edge $\ell$ is:\\
\begin{equation}
\label{eq:chooseY}
P(y|x,\ell)=\frac{1}{|Y(x,\ell)|}
\end{equation} 

The uniform assumption can be generalized to involve non-uniform probability, but in this work, since no model is used to decide the weights, uniform distribution is used.  

Given the Markov Chain and the weights described in Equation \ref{eq:chooseL} and \ref{eq:chooseY}, lazy walk, a variant of random walk that includes random stopping is used to traverse the knowledge graph. Lets denote $Q(x\xrightarrow{=d}z)$ as the probability that the random walk will end up in $z$ after $d$ steps, and lets denote:\\
\begin{equation}
\label{eq:Q1}
Q(x\xrightarrow{=0}z)=1
\end{equation} 
\begin{equation}
\label{eq:Q2}
Q(x\xrightarrow{=d}z)=\sum_{y}(\sum_{\ell}P(\ell|x).P(y|x,\ell)).Q(x\xrightarrow{=d-1}z)
\end{equation}  

Then, given a stopping probability of $\gamma$, the probability that the random walk will stop at $z$ after infinite steps is: 
\begin{equation}
\label{eq:Q3}
Q(z|x)=\gamma\sum_{d=1}^{\infty}(1-\gamma)^d Q(x\xrightarrow{=d}z)
\end{equation} 

As in \cite{Cohen2006}, instead of infinite steps, an approximation of $d=10$ will be used. Likewise, $\gamma$ is set to $0.5$.

Therefore, for every knowledge graph, $Q(z|x)$ will be generated to perform information retrieval. During querying, input is an initial distribution over all nodes $V(x)$, where $\sum_x V(x)=1$ . For example, if a question node $a$ is selected as input, then $V(a)=1$ whereas it is $0$ for the others. Similarity can then be calculated via $Q(z)=\sum_x Q(z|x)V(x)$ . The result will be filtered according to the intended output type and ranked.

\section{Case Study}
\label{sec:experiment}
Case study is performed on the Computer Science foundational course in the National University of Tainan. This course is based on the ``Foundations of Computer Science" textbook by Behrouz Forouzan \cite{forouzanTextBook}. 

The e-book is constructed such that they summarize some of the topics in the original printed book, in a way that meets the learning outcome for the course. There are altogether 11 chapters for the e-book, which covers topics ranging from computer architecture to programming languages and artificial intelligence. Domain experts construct the e-book in HTML format that covers both the description texts and questions for every chapter.  According to procedures described in Section \ref{sec:creatingFacts}, the topics and descriptions as well as their relationships are constructed. As this is a preliminary study, the intention of the case study is to determine performance given different construction of knowledge graphs, which at the same time, should not pose a huge overhead for domain experts. For controlled condition, no external knowledge base facts are used at the moment. Besides, the construction of most teaching materials doesn't have the luxury of a pre-made knowledge base that caters for its needs. Thus, after generating the facts, the database has about 1600 nodes and 14K triples. Given the small database, normal Matrix implementation of the random walk is used in this case study. For scalability, sampling approach can be used. 

\subsection{Evaluation on Different Construct of Knowledge Graph}
Knowledge graph is constructed given an initial specification of topic hierarchy, description sequence and topic-description link. Such knowledge graph may be too sparse for information retrieval. At the same time, to construct a dense graph, huge overhead is required. This is aggravated by the fact that more domain experts like teachers may not be familiar with knowledge graph construction. To alleviate this problem, one can use reasoners to generate more facts to enable denser graphs. In this work, FaCT++ is used for fact generation \cite{factplusplus} given the e-book ontologies provided by the domain experts, after which these facts are sent to the online database. FaCT++ is a reasoner for OWL DL, which is the ontology representation of the e-book. In this work, additional facts are generated pertaining to instance-class relationship, class-class relationship and property inverses. One can refer to \cite{AllemangBook2012} for more details about OWL DL and fact construction.

Evaluation is made between knowledge graphs with and without fact generation, as well as the inclusion of additional links from words.  Comparison is done using question query given  some inputs. Random input queries are constructed, and the ranked outputs are evaluation based on Mean Average Precision (MAP). The top 10 outputs are used for MAP calculation. As there are no benchmark or labeled data to compare to, a number of human evaluators are employed to determine the relevancy of the returned questions from the query. The MAP result is shown in Table \ref{tab:graph1}. It can be observed that with the generated facts via OWL reasoner, the query result will improve significantly. Such generation is automatic, thus, does not add to the e-book construction overhead that will be imposed on the domain expert. Although not shown, tests are also performed given larger step $d$ and lower stopping probability $\gamma$, but knowledge graph with the generated facts still fair significantly. This shows that the paths (that takes into account knowledge extracted from instance-class and class-class relationships) can effectively guide the random walk. Note that the low MAP score is due to some queries output having insufficient questions in the database. Given that such questions have been exhausted, the remaining output in the rank will likely be deemed irrelevant by human evaluators. In that case, one should have at least 10 questions in store for each possible queries.  

Word linkages (described in Section \ref{sec:additionalFacts}) will also contribute some improvements, though not significant, as currently, the database is too small to be judged. Despite that, word linkages are very important in that students can query with words as input. Another benefit is that, with such word linkages, nodes with no explicit linkages set by the domain expert can also be queried based on the texts contained inside.   

\begin{table}
	\scriptsize
	\caption{Ranked Information Retrieval Result under different Constructs of Knowledge Graph}
	\label{tab:graph1}
	\centering
	\begin{tabular}{|l|l|l|l|l|}
		\hline
		Knowledge & MAP & $\Delta MAP$ \\
		\hline
		e-book facts & 41.78  & - \\
		\hline
		e-book facts + word lingkages & 43.27 & 3.57 \%  \\
		\hline
		e-book facts + generated facts & 67.87 & 62.45 \%  \\
		\hline
		e-book facts + generated facts+word lingkages & 68.62 & 64.24 \%  \\
		\hline
	\end{tabular}
\end{table}
\normalsize

\subsection{e-Book Deployment}
\begin{figure}
	\begin{center}
		\includegraphics[angle=0, width=1\columnwidth]{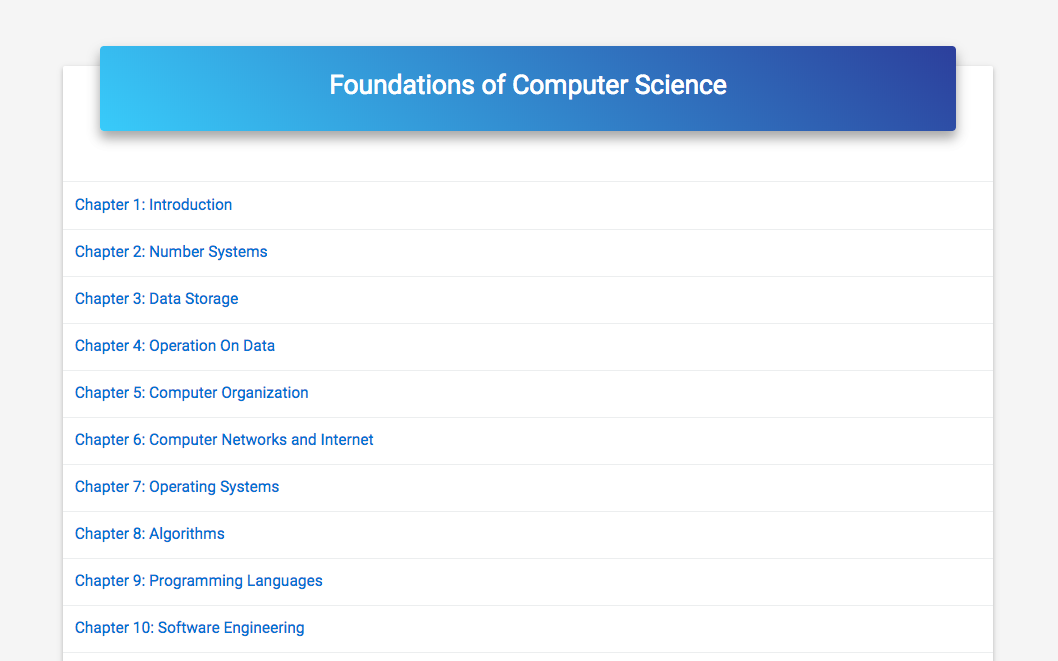}
	\end{center}
	\caption{Front Page of the e-book}
	\label{fig:ebookFront} 
\end{figure}

\begin{figure}
	\begin{center}
		\includegraphics[angle=0, width=1\columnwidth]{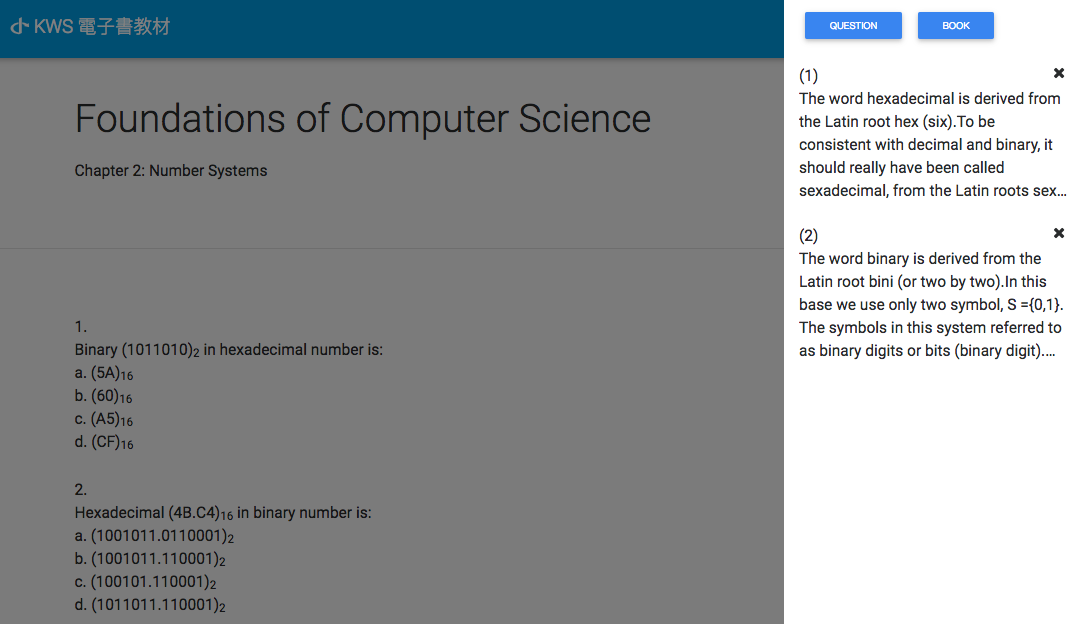}
	\end{center}
	\caption{Query Interface for the e-book}
	\label{fig:ebookRecord} 
\end{figure}

The e-book is set up as a website, which constantly updates its contents based on the database. As shown in Figure \ref{fig:architecture}, representation (website) is separated from fact generation modules, which eases the work for both domain expert and website designer as they don't have to consider the technicalities of their counterparts. The website is set up in such a way that students can easily scroll as well as interact with the texts. The front page of the e-book is shown in Figure \ref{fig:ebookFront}. Student may query for related questions or book texts given a set of descriptions they have recorded. For example, assuming the student have read about number system, and that he/she recorded the paragraph on hexadecimal number system and binary number system (where these two paragraphs do not mention about conversion between number systems). They can query for related questions, which will return results like questions for the conversion between hexadecimal and binary, ranked according to relevancy. This example query is shown in Figure \ref{fig:ebookRecord}.

\section{Conclusion}
\label{sec:conclusion}
With the advent of blended and personalized learning, technologies should be applied to reading for more efficient knowledge search, navigation and organization. Instead of solely replacing printed books, e-books should be augmented with such capabilities. This work is gearing towards this direction, where intelligent functionalities are endowed to the e-book without significant authoring overhead. This work assumes the book to be represented as a knowledge graph. To enable intelligent functionalities with the support of the knowledge graph, preliminary work on information retrieval within the e-book is done. Knowledge graph construction method is proposed, where additional facts are generated automatically from reasoners and word-linkages. Information retrieval is then realized through random process on the knowledge graph. Evaluation shows that the method of construction not only has little authoring overhead, but that relevant information can be retrieved. The e-book in this work is also been deployed as a website for students.


\section*{Acknowledgment}
This research is supported by the Ministry of Science and Technology of Taiwan (MOST 106-2811-E-024-003, MOST 106-3114-E-024-001, MOST 106-2221-E-024-019)



%
\bibliographystyle{IEEEtran}
\bibliography{reference}

\end{document}